\documentclass{aa}  

\usepackage{graphicx}
\usepackage{txfonts}
\begin{document}

   \title{Accretion and ejection at work in the Narrow Line Seyfert 1 galaxy \object{1H 0323+342}}

   \subtitle{A case of intermittent activity?}

   \author{Valentina Rosa
          \inst{1},
          Luigi Foschini \inst{2} 
          \and
          Stefano Ciroi \inst{1}
          }

   \institute{Department of Physics and Astronomy, University of Padova, vicolo dell'Osservatorio 3, 35122 Padova, Italy
         \and
            Brera Astronomical Observatory, National Institute of Astrophysics (INAF), via Bianchi 46, 23807 Merate, Italy
             }
  
 \abstract
   {We present a comprehensive investigation into the properties of \object{1H 0323+342}, a prominent jetted narrow-line Seyfert 1 galaxy.}
   {The primary objective is to understand the interplay between the relativistic jet, the hot corona, and the accretion disk around the supermassive black hole.}
   {This study spans the years 2006 to 2023, incorporating a rich dataset with 172 \emph{Swift} observations, including the optical, UV, and X-ray bands, integrated with \emph{Fermi} Large Area Telescope (LAT) observations. Spectral analysis was conducted on the X-ray observations using the XSPEC software, and the results were compared with optical, UV, and $\gamma$-ray flux measurements and photon index values.}
   {Our key findings include the identification of three distinct zones in the X-ray photon index–flux plot, characterized by high flux and a hard photon index (zone 1), high flux and a soft photon index (zone 2), and low flux and a soft photon index (zone 3). Before $\sim 2017$, 1H~$0323+342$ moved back and forth between zones 1 and 2; after that epoch, it transitioned to zones 2 and 3. Correspondingly, we observed a decreasing jet activity in the \emph{Fermi}/LAT data and a reduction in the accretion rate in optical/UV data from \emph{Swift}/UVOT.}
   {We interpret these observations in the framework of an intermittent jet scenario, driven by radiation-pressure instability in the accretion disk.}
   

\titlerunning{Accretion and ejection at work in the Narrow Line Seyfert 1 galaxy \object{1H 0323+342}}
\authorrunning{V. Rosa et al.}
\maketitle

\section{Introduction} \label{sec:intro}
Narrow-line Seyfert 1 galaxies (NLS1) are a distinct subset of active galactic nuclei (AGNs) that exhibit unique optical features, including broad Balmer lines with full widths at half maximum $\lesssim 2000$~km~s$^{-1}$, weak [OIII] emission lines with a flux ratio of [OIII] to H$\beta < 3$, and strong \ion{Fe}{II} bumps \citep{osterbrock1985spectra,GOODRICH1989}. 

In addition to these optical properties, NLS1s exhibit bright X-ray emission with steeper spectra compared to their broad-line counterparts \citep{Leighly1999}. These galaxies typically host supermassive black holes with relatively low masses, ranging from $10^6$ to $10^8 M_{\odot}$ \citep{PETERSON2011,PETERSON2018}, and accrete close to the Eddington limit. This suggests they are young AGNs that may evolve into more luminous quasars \citep{grupe2004mbh}. 

The discovery of high-energy $\gamma$-ray emission from NLS1s with the \emph{Fermi} Large Area Telescope (LAT) in 2008 renewed interest in this class of AGNs \citep{abdo2009radio}; see also \citet{FOSCHINI2012,FOSCHINI2020} for comprehensive reviews.

\object{1H 0323+342} is the closest $\gamma$-ray-emitting NLS1 ($z = 0.063$). Initially classified as a Seyfert 1 galaxy in the HEAO-1 X-ray survey, it showed strong \ion{Fe}{II} lines and weak forbidden lines \citep{wood1984heao,remillard1993twenty}. Subsequent observations, particularly in the radio band, revealed its hybrid nature, making it a prototype for a new class of AGNs that share characteristics of both NLS1s and blazars \citep{Zhou2007}. 

\emph{Fermi}/LAT observations confirmed the presence of a relativistic jet, which contributes to the nonthermal emission observed across its broadband spectral energy distribution (SED). The SED shows two distinct peaks in the radio–infrared and GeV $\gamma$-ray bands, along with variations in the X-ray emission, all of which indicates a complex interplay between the emission components \citep{abdo2009radio}.

Given its proximity to Earth, \object{1H 0323+342} is an ideal target for detailed study, allowing exploration across a broad range of wavelengths with higher angular resolution. Over the past decade, various observational campaigns have been conducted at radio frequencies \citep{wajima2014short,FUHRMANN2016,doi2018recollimation,HADA2018,DOI2020}, optical wavelengths \citep{ITOH2014,LEONTAV2014,wang2016reverberation,WANG2017,OLGUIN2020,TURNER2022}, X-rays \citep{YAO2015,GHOSH2018,BERTON2019,MUNDO2020}, and in multiwavelength studies \citep{TIBOLLA2013, paliya2014peculiar,landt2017black,YAO2023,LUASHVILI2023}.

A summary of the physical properties of \object{1H 0323+342} is presented in \citet{foschini2019mapping}. Notably, the X-ray spectrum of \object{1H 0323+342} is generally dominated by a power law with $\Gamma \sim 2.2$, likely due to the thermal Comptonization of the hot corona surrounding the accretion disk. Occasionally, however, a harder component ($\Gamma \sim 1.4$) emerges at a few keV (see, in particular the left panel in Fig.~1, in \citealt{FOSCHINI2012}), which suggests enhanced activity in the relativistic jet \citep{FOSCHINI2009, FOSCHINI2012, TIBOLLA2013, paliya2014peculiar}. Therefore, \object{1H 0323+342} provides an ideal case study for examining the interplay between the accretion disk and the relativistic jet. To this end, we (re)analyzed all available \emph{Swift} observations in the archive to further investigate this source’s properties.

\section{Data analysis} \label{sec:style}
Data from the \emph{Neil Gehrels Swift Observatory} \citep{gehrels2004swift} were obtained from the High-Energy Astrophysics Science Archive Research Center (HEASARC\footnote{\url{https://heasarc.gsfc.nasa.gov/}}). All \emph{Swift} observations within $10'$ of the coordinates of \object{1H 0323+342} were selected, resulting in 173 observations. The dataset includes X-Ray Telescope (XRT; \citealt{XRT}) and Ultra-Violet/Optical Telescope (UVOT; \citealt{UVOT}) data collected from 2006 to 2023. One observation (ID: 03111698008, 2022 November 11) was excluded because it contained only window-timing data recorded during slew and settling.

Data reduction was performed using \texttt{HEASOFT 6.31.1} with calibration files from \texttt{CALDB} (updated on 2024 May 13). Standard procedures with default parameters were applied, as described in "The Swift XRT Data Reduction Guide" (Version 1.2, April 2005). The X-ray spectra were rebinned to ensure a minimum of 20 counts per bin, allowing the use of $\chi^2$ statistics.  

X-ray spectra were analyzed using the \texttt{xspec} software package. To study the accretion-ejection evolution, we applied simple models (power law and broken power law) absorbed by the Galactic column (\texttt{tbabs} model with $N_{\rm H}=1.17\times 10^{21}$~cm$^{-2}$; \citealt{GALACTICNH}). Model selection was performed using an F-test, with the broken power law model adopted when the probability value was below $4.0 \times 10^{-3}$. Eleven observations were best fitted by the broken power law model. Unabsorbed fluxes were computed using the \texttt{cflux} convolution model. In five cases (2015 December 3, 2018 September 13, 2019 October 18, 2022 August 27, and 2023 February 11), the inclusion of the \texttt{cflux} model resulted in insufficient counts for $\chi^2$ statistics. Therefore, $\chi^2$ statistics were employed only for model selection, and the C-statistic was used to determine flux values for these observations, applied to spectra without a minimum count threshold per bin. Unless explicitly stated otherwise, all reported fluxes are intrinsic. Confidence intervals (90\%) for the fitted parameters were estimated using the \texttt{error} command. Features such as the narrow Fe K$\alpha$ line, previously reported by \cite{abdo2009radio} and \cite{paliya2014peculiar}, were not modeled, as their analysis would require longer exposure times and lies beyond the scope of this work.

UVOT magnitudes were extracted from a circular region with a $5''$ radius, while the background was measured from an annular region with inner and outer radii of $7''$ and $30''$, respectively. Extinction corrections were applied to the UVOT magnitudes using the visual absorption $A_{\rm V} = 0.578$ from the Galactic dust reddening map \citep{schlafly2011measuring}, following the extinction laws of \cite{cardelli1989relationship}. The corrected magnitudes were converted into flux densities using zero-point values provided by \texttt{CALDB}, and two-point spectral indices were computed as:

\begin{equation} \centering \alpha_{\rm UV} = -\frac{\log \frac{S_{\rm UVW1}}{S_{\rm UVW2}}}{\log \frac{\nu_{\rm UVW1}}{\nu_{\rm UVW2}}}, \quad
\alpha_{\rm opt} = -\frac{\log \frac{S_{\rm V}}{S_{\rm U}}}{\log \frac{\nu_{\rm V}}{\nu_{\rm U}}},
\label{alfao} \end{equation}

\noindent where $\nu_{\rm V} = 5.56 \times 10^{14}$~Hz, $\nu_{\rm U} = 8.57 \times 10^{14}$~Hz, $\nu_{\rm UVW1} = 1.16 \times 10^{15}$~Hz, and $\nu_{\rm UVW2} = 1.48 \times 10^{15}$Hz. For each observation, the derived values include: X-ray fluxes, photon indices, UV and optical flux densities, as well as spectral indices. A complete set of results is provided in the Appendix.

\section{Results}

Figure~\ref{fig:Graph Zone} presents the X-ray flux in the $0.3$--$10$~keV energy band as a function of the photon index.
The distribution of data points suggests setting two red lines at $F_{\rm 0.3-10, keV} \sim 10^{-11}$ erg cm$^{-2}$ s$^{-1}$ and $\Gamma \sim 1.9$. These lines divide the diagram into four “zones”:

\begin{figure*}[!ht]
    \centering
    \includegraphics[width=0.9\textwidth]{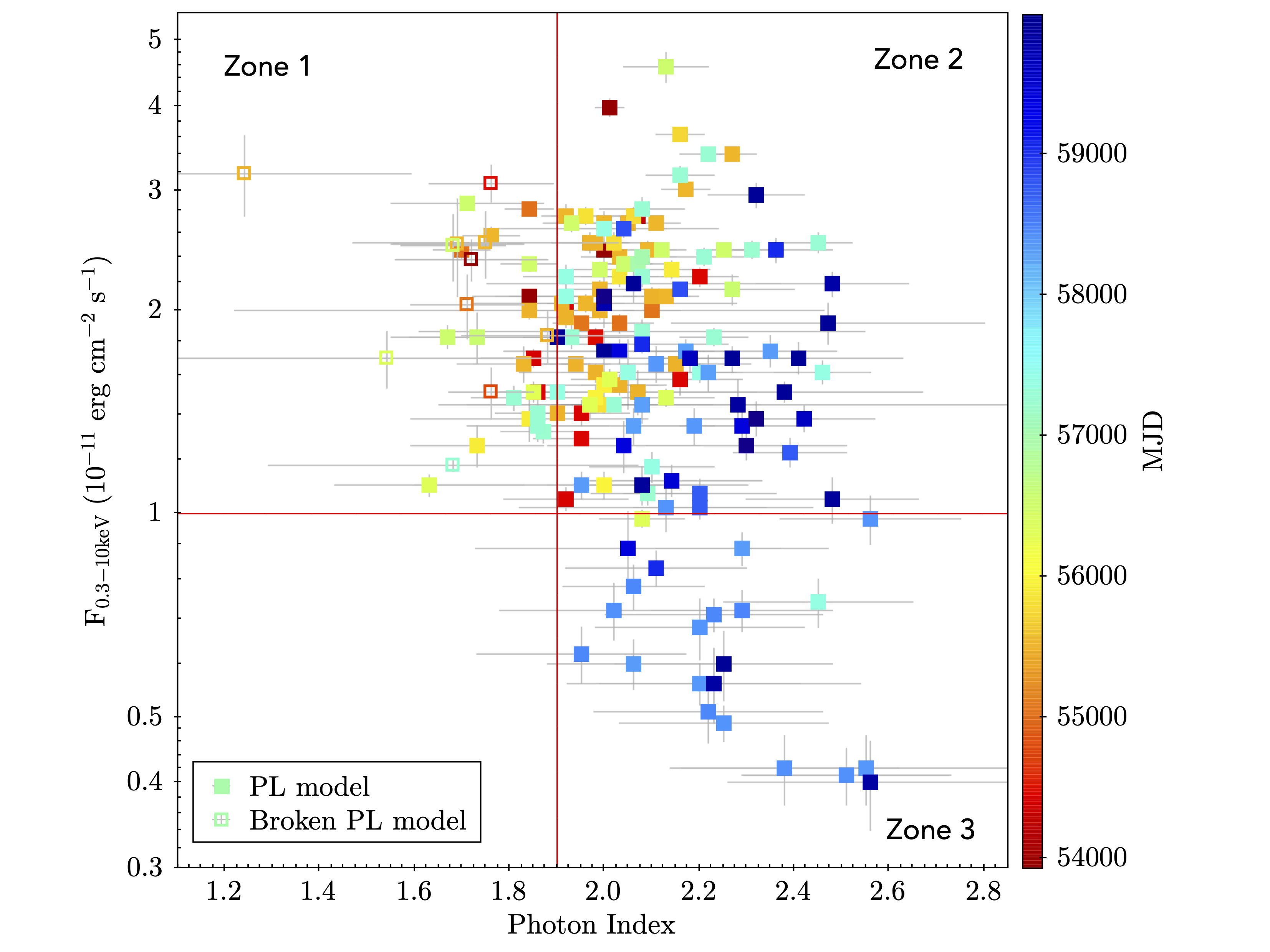}
    \caption{
    Intrinsic X-ray flux in the 0.3–10 keV energy band as a function of the photon index. For observations modeled with a broken power law, the hard photon index and total flux are plotted. The vertical red line is set at a photon index of $\Gamma \sim 1.9$, and the horizontal red line at a flux of $F_{\rm 0.3-10\, keV} \sim 10^{-11}$ erg cm$^{-2}$ s$^{-1}$. The plot is divided into three “zones”: zone 1 (top left) represents observations best modeled by a broken power law or a power law with a hard photon index; zone 2 (top right) includes observations modeled by power laws with softer photon indices and higher flux values; and zone 3 (bottom right) corresponds to observations modeled by power laws with soft photon indices and lower fluxes. The color gradient on the right represents the modified Julian date (MJD) of the observations, illustrating the temporal evolution of the source’s behavior.
    \label{fig:Graph Zone}}
\end{figure*}

\begin{itemize}
\begin{small}
    \item Zone 1 (top left): $\Gamma$ $ \lesssim$ 1.9 \& $F_{\rm 0.3-10\, keV}$ $ \gtrsim$  $10^{-11}$~erg~cm$^{-2}$~s$^{-1}$;
    \item Zone 2 (top right):  $\Gamma$ $ \gtrsim$ 1.9 \& $F_{\rm 0.3-10\, keV}$ $ \gtrsim$  $10^{-11}$~erg~cm$^{-2}$~s$^{-1}$;
     \item Zone 3 (bottom right): $\Gamma$ $ \gtrsim$ 1.9 \& $F_{\rm 0.3-10\, keV}$ $ \lesssim$  $10^{-11}$~erg~cm$^{-2}$~s$^{-1}$;
     \item Zone 4 (bottom left): $\Gamma$ $ \lesssim$ 1.9 \& $F_{\rm 0.3-10\, keV}$ $ \lesssim$  $10^{-11}$~erg~cm$^{-2}$~s$^{-1}$, zone of avoidance, no X-ray data were found in this zone.
\end{small}    
\end{itemize}

To categorize the observations into spectral states, we initially fitted all spectra with a simple power law model, identifying an average photon index of 1.9. This threshold ($\Gamma \sim 1.9$ ) aligns with the typical photon index for unsaturated Comptonization \citep{RYBLIG}, the process governing radiative emission in the corona of an accretion disk (e.g., \citealt{HAARDT1991,HAARDT1993}). As a result, we identify zones 2 and 3 as primarily dominated by coronal emission, whereas zone 1, which exhibits higher fluxes and harder photon indices, is likely dominated by relativistic jet emission.

Observations with a photon index below this threshold were categorized as belonging to zone 1. Following this initial classification, we refined the spectral fitting by selecting either a simple power law or a broken power law model, choosing the one that provided the best fit. For spectra best described by a broken power law, we used the hard photon index (corresponding to the high-energy component) for classification.

After classifying the three zones, we combined the spectra corresponding to each zone using the \texttt{addspec} tool, weighting each observation by its duration. Initially, individual spectra were not binned with a minimum count threshold. The co-added spectra were subsequently re-binned to ensure a minimum of 30 counts per bin. For zone 1, the broken power law model provided the best fit, whereas for zone 2, the addition of a broken power law component did not significantly improve the fit over a simple power law. Instead, including a \texttt{zbbody} component to model the soft excess improved the fit. For zone 3, the simple power law model was sufficient to describe the spectra. While more complex components, particularly in the soft X-ray band, could further refine the fits for zones 1 and 2, the main focus of this work is on the jet emission, which primarily contributes to the hard X-rays. The resulting spectral fits are summarized in Table~\ref{tab:3 ZONE FITTING}.

\begin{table*} 
\caption{Spectral fits of the co-added spectra for the three zones. \label{tab:3 ZONE FITTING}} \centering 
\begin{tabular}{c c c c c c c c} 
\hline
\hline 
Zone & Model & $\Gamma_1$ & $F_1$ & E$_\mathrm{break}$ & $\Gamma_2$ & $F_2$ & $\chi^2$\\ 
(1) & (2) & (3) & (4) & (5) & (6) & (7) & (8) \\ 
\hline
 1 & \texttt{bkn} & 1.96 $\pm$ 0.03 & 1.63 $\pm$ 0.10 & 2.47 $\pm$ 0.11 & 1.74 $\pm$ 0.05 & 0.68 $\pm$ 0.08 & 410.54/288 \\ 
 2 & \texttt{pow} & 2.07 $\pm$ 0.02 & 1.93 $\pm$ 0.04 & \dots & \dots & \dots & 538.94/377 \\ 
 3 & \texttt{pow} & 2.21 $\pm$ 0.05 & 0.89 $\pm$ 0.07 & \dots & \dots & \dots & 130.01/124 \\ 
 \hline 
 \end{tabular}

\textbf{Notes}: Column description: (1) Zone as shown in Fig.~\ref{fig:Graph Zone}; (2) Best-fit model (\texttt{bkn} = broken power law; \texttt{pow} = power law); (3) $\Gamma_1$, photon index for the simple power law model or soft photon index for the broken power law model; (4) $F_1$, flux in the 0.3–10 keV band for the power law model, or soft component flux for the broken power law model [$10^{-11}$ erg cm$^{-2}$ s$^{-1}$]; (5) $E_{\rm break}$, break energy for the broken power law model [keV]; (6) $\Gamma_2$, hard photon index of the broken power law model; (7) $F_2$, flux of the hard component in the broken power law model [$10^{-11}$ erg cm$^{-2}$ s$^{-1}$]; (8) $\chi^2$/dof. \end{table*}

To investigate the contributions of the accretion disk and the jet, we examined observations at wavelengths where each component is expected to dominate. Previous SED studies suggest that the optical/UV emission primarily originates from the accretion disk \citep{abdo2009radio, FOSCHINI2012}.

Although the most significant variations occur in the X-ray band, our analysis of UVOT data also reveals variability in both flux and spectral index. To compare the three zones across different energy bands, we computed the average fluxes and spectral indices for each zone, as summarized in Table~\ref{tab: spectral}. Zones 1 and 2 have consistent values within the uncertainties, whereas zone 3 has lower fluxes and softer spectral indices. A complete list of results is provided in the Appendix.

\begin{table}
\caption{Average spectral indexes, optical and UV fluxes for each zone}.   \label{tab: spectral} 
\centering 
\begin{tabular}{c c c c c c c c} 
\hline\hline 
Zone & $\alpha_{\mathrm{UV}}$ & $\alpha_{\mathrm{opt}}$ & $\mathrm{F}_{\mathrm{UV}}$ & $\mathrm{F}_{\mathrm{opt}}$ \\ 
 (1) &  (2) & (3) & (4) & (5) \\
\hline 
1 & -0.20 $\pm$ 0.08 & 0.67 $\pm$ 0.05 & 2.43 $\pm$ 0.03 & 2.06$\pm$ 0.04 \\
2 & -0.10 $\pm$ 0.08 & 0.64 $\pm$ 0.04 & 2.39 $\pm$ 0.04 & 2.00 $\pm$ 0.03\\
3 &  0.02 $\pm$ 0.09 & 0.86 $\pm$ 0.07 & 2.14 $\pm$ 0.05 & 1.83 $\pm$ 0.06  \\
\hline 
\end{tabular}

\textbf{Notes}: Column description: (1) Zone from Fig.~\ref{fig:Graph Zone}; (2) average ultraviolet spectral index from Eq.~(\ref{alfao}); (3) average optical spectral index from Eq.~(\ref{alfao}); (4) average ultraviolet flux [$10^{-11}$~erg~cm$^{-2}$~s$^{-1}$]; (5) average optical flux [$10^{-11}$~erg~cm$^{-2}$~s$^{-1}$].
\end{table}

\begin{figure*}[!ht]
    \centering
    \includegraphics[width=0.9\textwidth]{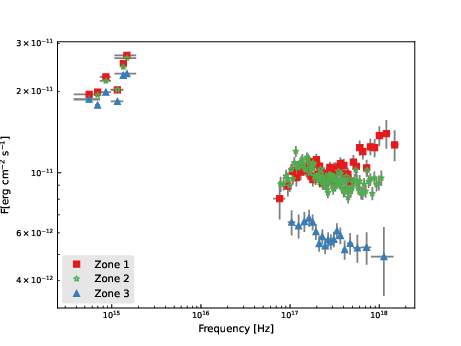}
    \caption{
    SED of \emph{Swift} observations of 1H0323+342. Left: UVOT data. Right: XRT data. The plot shows unabsorbed flux (erg cm$^{-2}$ s$^{-1}$) on a logarithmic scale as a function of frequency (Hz), with the three distinct zones indicated: zone 1 as red squares, zone 2 as green stars, and zone 3 as blue triangles.}
\label{fig:SED}
\end{figure*}

Additionally, we constructed a SED for \object{1H 0323+342}, shown in Fig.~\ref{fig:SED}, covering all \textit{Swift} frequency ranges. This confirms our earlier findings: while variations in UVOT frequencies are less pronounced than in XRT, zone 3 consistently exhibits lower flux levels.

\begin{figure}[h!]
\centering
  \includegraphics[width=0.5\textwidth]{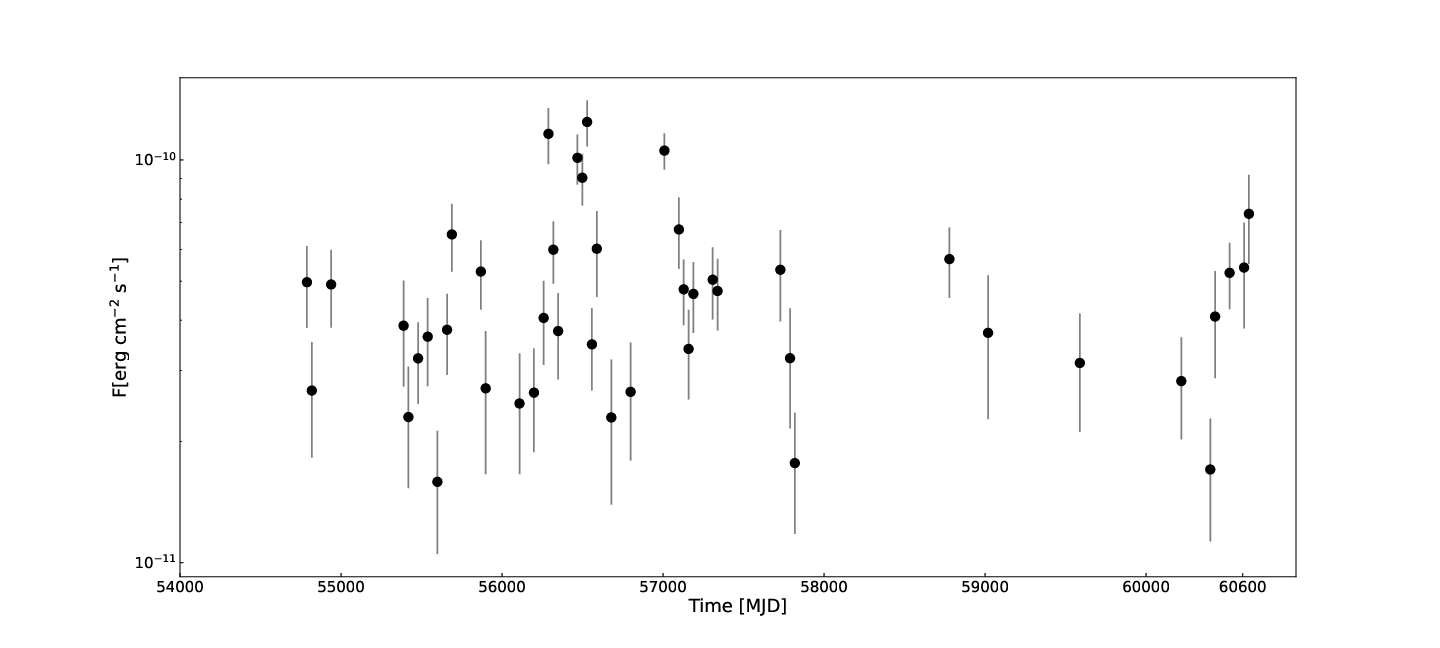}
  \includegraphics[width=0.5\textwidth]{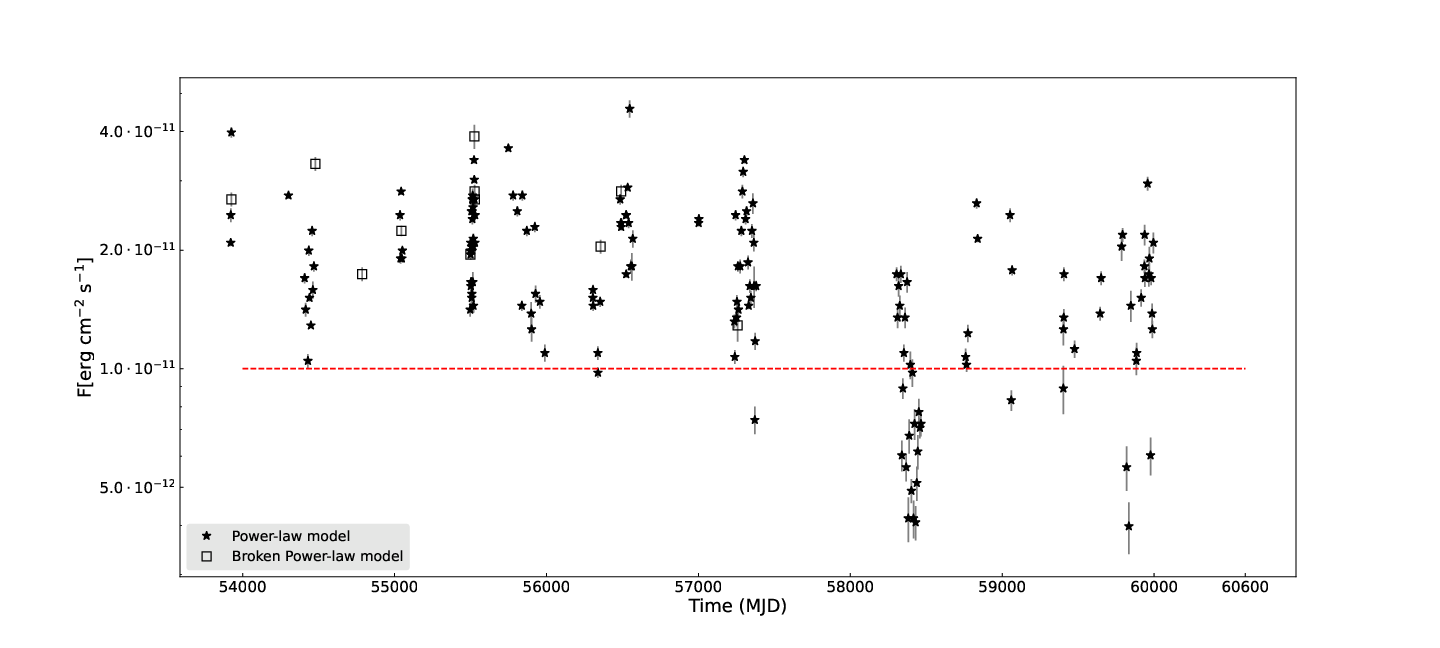}
  \caption{(\emph{top panel}) 
\emph{Fermi}/LAT light curve in the $0.1-100$~GeV energy range, with a clear absence of detections beginning around 2017 September 4 (MJD 58000). (\emph{bottom panel})  \emph{Swift}/XRT light curve in the $0.3-10$~keV energy band, with the dashed red line set at flux, $F_{\rm 0.3-10, keV} \sim 10^{-11}$ erg cm$^{-2}$ s$^{-1}$. We see a decrease in flux values starting from 2017 (MJD 57500-58000) }
  \label{fig: lightcurves}
\end{figure}

\begin{figure}[h!]
\centering
  \includegraphics[width=0.5\textwidth]{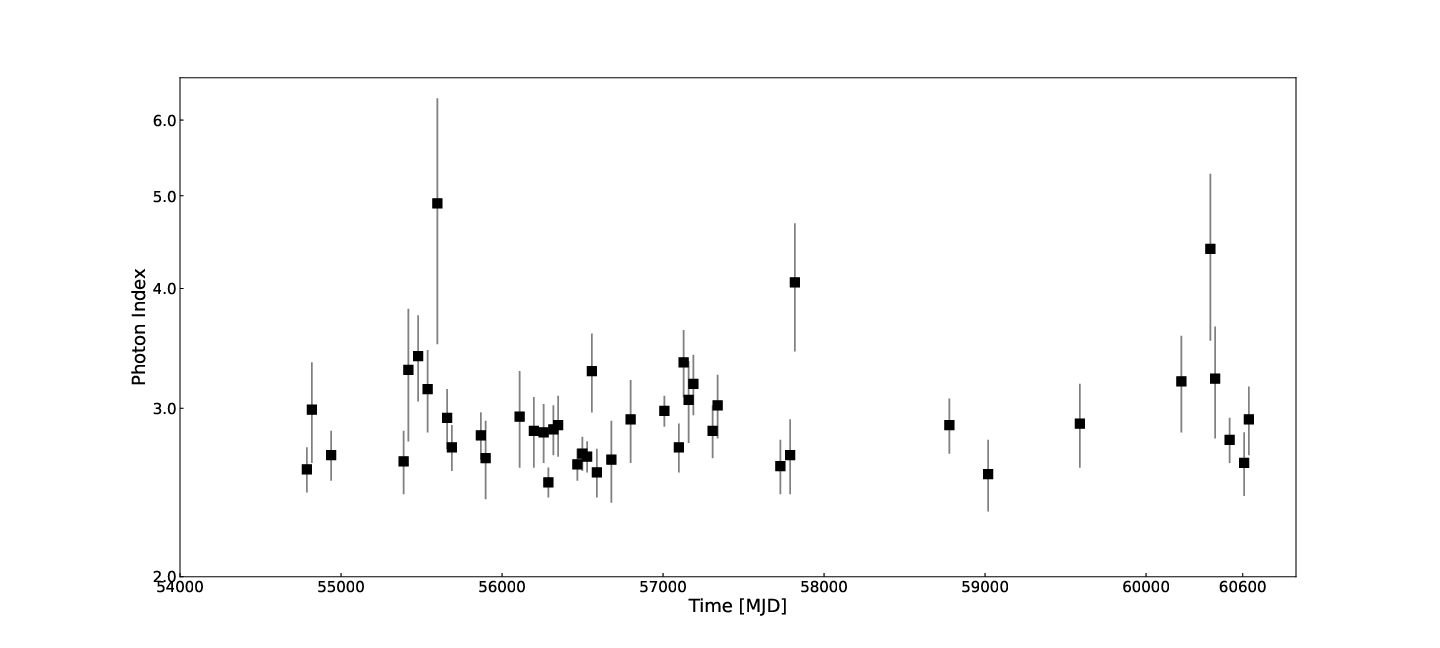}\\
  \includegraphics[width=0.5\textwidth]{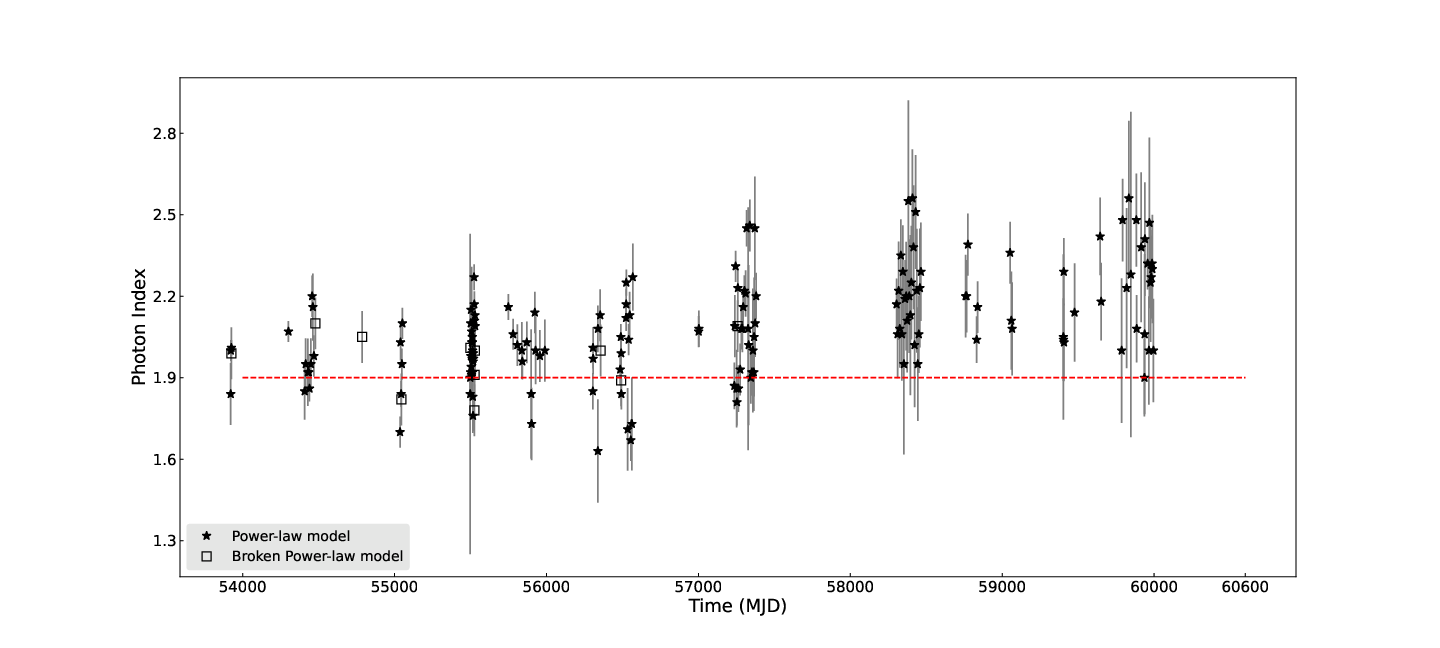} 
  \caption{(\emph{top panel})  Evolution of the \emph{Fermi}/LAT photon index from 2007 to 2024 ($\sim$ MJDs 54000-60600). (\emph{bottom panel}) Changes in the photon index in the same time period with the dashed red line set at photon index $\Gamma$ = 1.9.}
  \label{fig:photons}
\end{figure}

Jet emission dominates the high-energy $\gamma$-ray band, requiring data from \textit{Fermi}/LAT. The \textit{Fermi} LAT Light Curve Repository (LCR\footnote{\url{https://fermi.gsfc.nasa.gov/ssc/data/access/lat/LightCurveRepository/index.html}}, \citealt{FERMILCR}) provides publicly available light curves in the $0.1-100$~GeV band, at intervals of 3 days, 1 week, and 1 month, with fixed or free photon index. We retrieved the light curve of \object{1H 0323+342} by selecting three-day time bin and free $\Gamma$. We considered only values with likelihood test statistic $TS \gtrsim 10$, equivalent to $\gtrsim 3\sigma$ detection \citep{MATTOX1996}. Data are shown in the top panels of Fig.~\ref{fig: lightcurves} \ref{fig:photons}. 

We compared the \emph{Fermi}/LAT curves (Fig.~\ref{fig: lightcurves}) with the \emph{Swift}/XRT ones. We noted that the lack of $\gamma$-ray detection corresponds to a lower X-ray flux, and steeper spectra. Interestingly, these data refer to the zone 3.  

\begin{figure}
    \centering
    \includegraphics[width=0.5\textwidth]{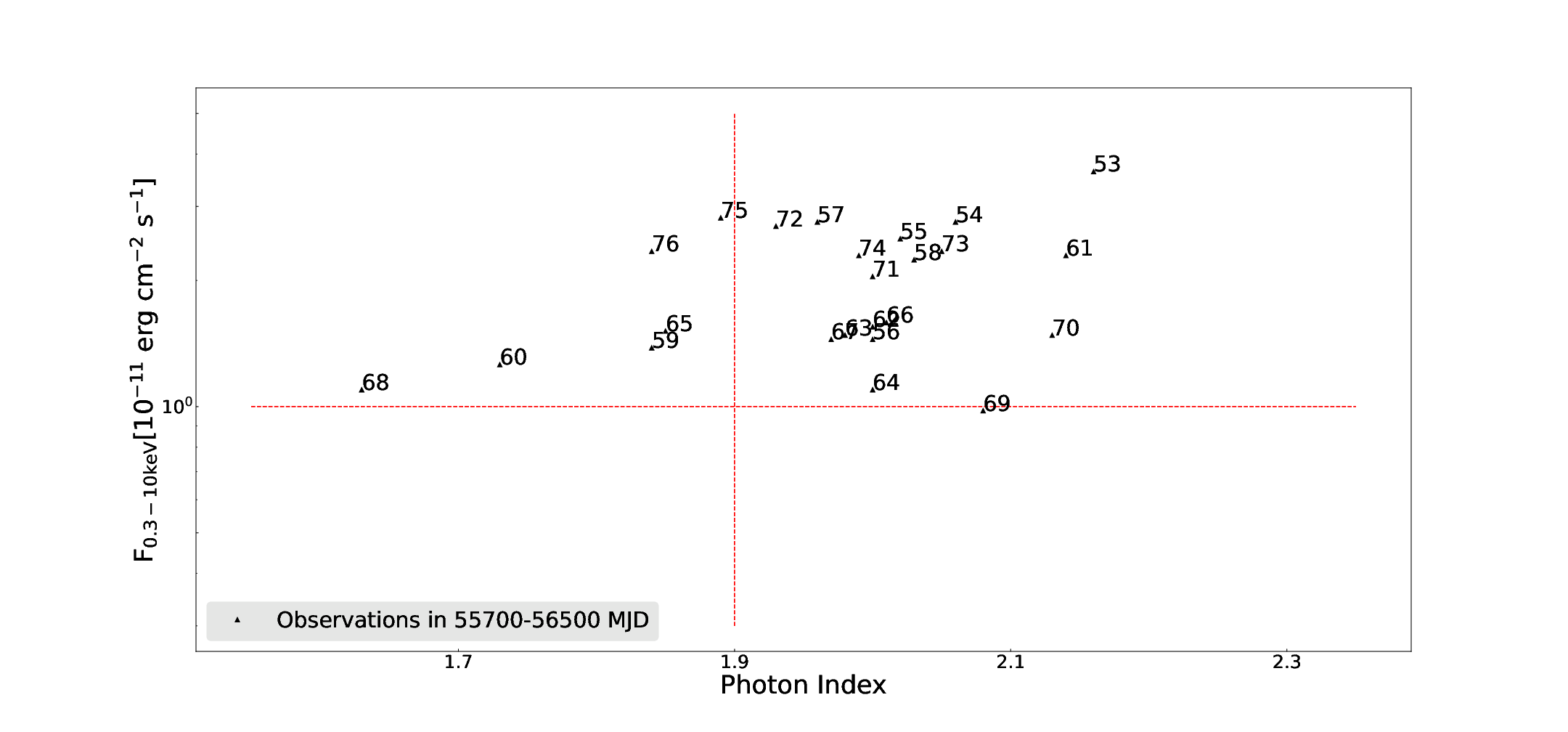}
    \caption{Photon index vs. $0.3-10$~keV flux before and during the first 2013 $\gamma$-ray outburst. Observations span from 2011 May 19 (MJD 55700) to 2013 July 27 (MJD 56500), with numbers indicating the chronological sequence. The movement of data points between zone 1 and zone 2 highlights the evolution of the X-ray spectral state during jet activity.}
    \label{fig: gammarayburst}
\end{figure}

In summary, significant jet activity was observed between 2006 and 2017, as indicated by the X-ray data fluctuating between zones 1 and 2, accompanied by prominent high-energy $\gamma$-ray emission, including a notable outburst in 2013 \citep{CARPENTER2013, paliya2014peculiar, PALIYA2015}. As shown in Figure~\ref{fig: gammarayburst}, the data points move between zone 1 and zone 2 during the jet activity, illustrating the spectral evolution associated with the $\gamma$-ray outburst. This transition provides insights into the interplay between the relativistic jet and the corona, with higher flux states accompanied by spectral hardening, a characteristic signature of jet-dominated emission.

Following 2017, the source shifted to zone 3, marked by lower X-ray flux and steeper spectra, although occasional flux increases were detected, consistently associated with soft photon indices (i.e., the source alternated between zones 3 and 2). A similar trend was observed in the optical/UV bands, and $\gamma$-ray detections became progressively less frequent.

\section{Discussion}
Being the closest NLS1, \object{1H 0323+342} is a bright source with strong X-ray emission, extensively observed by various satellites, particularly after the discovery of high-energy $\gamma$-ray emission from its relativistic jet. This makes it an excellent target for investigating jet-disk coupling and emission mechanisms. Many studies have focused on the corona, the potential soft excess, the Fe K$\alpha$ emission line, the spin of the central black hole, and SED modeling (e.g., \citealt{YAO2015,KYNOCH2018,GHOSH2018,MUNDO2020,LUASHVILI2023}). 

For example, \cite{LUASHVILI2023} modeled different states of \object{1H 0323+342} between 2008 and 2015, but their analysis did not include zone 3, which occurred after 2017. Their SED modeling confirms our findings, showing changes in jet emission primarily caused by variations in the electron distribution, while the surrounding photon field remained nearly constant. In their work, the source is observed moving back and forth between zones 1 and 2, consistent with our classification scheme.

In other studies, \object{1H 0323+342} has been analyzed as part of larger samples, with its X-ray spectral characteristics examined in a broader context (e.g., \citealt{WALTON2013,PALIYA2015,ISO2016,BERTON2019,WADDELL2020,DAMMANDO2020}). X-ray spectral variability has been widely reported, often showing spectral hardening during jet activity (\citealt{FOSCHINI2009,FOSCHINI2012,TIBOLLA2013,paliya2014peculiar,PALIYA2015}).

\emph{Swift} monitoring over the past two decades has been particularly valuable due to the simultaneous availability of optical-to-X-ray data (e.g., \citealt{TIBOLLA2013,paliya2014peculiar,PALIYA2015,YAO2015,DAMMANDO2020}). For instance, \cite{YAO2023} analyzed all \emph{Swift} observations from 2006 to 2021 and identified a strong correlation between the $\alpha_{\rm ox}$ parameter, defined as:

\begin{equation} \alpha_{\rm ox}=-0.384\log \frac{f_{\rm 2500\AA}}{f_{\rm 2, keV}},
\end{equation}

\noindent and the X-ray flux in the $0.3-2$~keV and $2-10$~keV bands. They observed that the broad-band spectrum becomes bluer during brightening episodes. \cite{YAO2015} attributed long-term UV/X-ray variability to the viscous timescale of an optically thick accretion disk \citep{CZERNY2006}, while short-term variability was linked to changes in the mass density and temperature of the X-ray corona. When the jet is inactive, it contributes less than $\sim 5$\% of the total optical-to-X-ray flux \citep{YAO2023}.

Our optical, UV, and X-ray data from \emph{Swift} observations reveal three distinct zones of activity for \object{1H 0323+342}, with a threshold X-ray flux of $\sim 10^{-11}$ erg cm$^{-2}$ s$^{-1}$ and a photon index of $\sim 1.9$. The novelty of this work is the identification of zone 3. The optical and UV fluxes exhibit similar behavior within zones 1+2 and 3, although with smaller amplitude variations compared to the X-ray band. This is consistent with a minimal jet contribution in these bands, as inferred from the low polarization percentage (1--3\%, \citealt{ITOH2014}). The jet activity observed in the X-ray band aligns with the $\gamma$-ray light curve from Fermi/LAT.

Fermi/LAT observations began in August 2008, but \object{1H 0323+342} was not detected during the first three months of operations \citep{abdo2009bright}. After one year, however, it was identified as one of three newly detected RL-NLS1 sources with a detection significance of approximately $5\sigma$ \citep{abdo2009radio}. Multiwavelength light curves from \cite{paliya2014peculiar} also revealed a radio outburst at 15 GHz in 2009, following a period of low X-ray flux ($\sim 10^{-12}$ ergcm$^{-2}$~s$^{-1}$) reported by \cite{Zhou2007}, indicating near-inactive jet conditions before $\gamma$-ray detection. 

Although more recent \emph{Swift} data are unavailable, Fermi/LAT observations indicate renewed jet activity in recent months, following a period of inactivity lasting approximately 7--8 years. These observations align with the interpretation of intermittent jet activity driven by radiation-pressure instability in the accretion disk \citep{czerny2009accretion}.

According to \cite{czerny2009accretion}, such instabilities occur in objects with high accretion rates, including NLS1s like \object{1H 0323+342}, which are part of the same population as compact symmetric objects  viewed at different angles \citep{berton2016compact}. These instabilities require a dimensionless accretion rate $\dot{m} \gtrsim 0.025$, and the time interval between active phases is given by the equation:

\begin{equation}
\footnotesize
\log T~[{\rm y}] \sim 1.25 \log(\nu 1L_{\rm 5GHz}) - 0.38 \log \left(\frac{\alpha}{0.02}\right) + 1.25 \log K_{5} - 53.6.
\label{czerny} \end{equation}

\noindent where $L_{\rm 5GHz}$ [erg~s$^{-1}$] is the monochromatic radio luminosity at 5~GHz, $K_{5}$ is the bolometric correction factor at the same frequency, and $\alpha$ is the viscosity of the accretion disk. 

For \object{1H 0323+342}, we redefined the X-ray threshold flux by separating the contributions from the jet and the corona. Based on observations modeled with a broken power law, the soft component is attributed to the corona, while the hard component arises from the jet. Our analysis (see Table A.1.) indicates that the jet contributes between 20\% and 77\% of the total flux, with an average of 42\%. Consequently, we attributed 58\% of the threshold flux ($\sim 5.8 \times 10^{-12}$ erg cm$^{-2}$ s$^{-1}$) to the corona. Using photon index value $\Gamma$=1.9, we scaled this flux from the 0.3--10 keV band to the 2--10 keV band and estimate the bolometric luminosity following the \cite{netzer2019bolometric} law. The resulting 2--10 keV luminosity is $\sim 2.5 \times 10^{43}$ erg s$^{-1}$, with a bolometric correction factor $K_{2-10 \rm keV} \sim 18.4$, yielding a bolometric luminosity of $4.6 \times 10^{44}$ erg s$^{-1}$.
Given a black hole mass of $\sim 2 \times 10^{7} M_{\odot}$ \citep{landt2017black}, the Eddington luminosity is $2.6 \times 10^{45}$ erg s$^{-1}$, resulting in a dimensionless accretion rate of:

\begin{equation} \dot{m} = \frac{L}{L_{\rm Edd}} \sim \frac{4.6 \times 10^{44}}{2.6 \times 10^{45}} \sim 0.18,
\end{equation}

which significantly exceeds the threshold for radiation-pressure instability \citep{czerny2009accretion}.

To calculate the recurrence time, we refer to the radio luminosity at 5 GHz. \cite{angelakis2015radio} report a stable average radio flux of 0.4 Jy at 4.85 GHz during 2009–2014, with minimal variations during flares,  where the Doppler factor reaches values up to 3 only. Therefore, assuming no significant Doppler boosting, the corresponding radio luminosity is $\sim 1.7 \times 10^{41}$ erg s$^{-1}$, with a bolometric correction factor $\log K_{5} \sim 3.4$. Substituting these values into Eq.~(\ref{czerny}), we estimate a recurrence time of $\sim 154$ years.

However, as \cite{czerny2009accretion} noted, higher viscosity or accretion rates near the Eddington limit can reduce this timescale by an order of magnitude, consistent with the observed 7–8 year interval. The elevated magnetic field inferred from the SED model of \cite{abdo2009radio} suggests high viscosity, further supporting this adjustment.

While these calculations provide reasonable estimates, caution is warranted due to uncertainties, such as the limited temporal sampling, challenges in separating jet and corona contributions in the X-ray band, and assumptions about viscosity. Nonetheless, the theory of radiation-pressure instability remains a plausible explanation for the observed intermittent jet behavior, whilst requiring fine-tuning and detailed modeling. A targeted multiwavelength monitoring campaign could provide further insights into this phenomenon.

\section{Comparison with the intensity-hardness plot of stellar-mass black holes}

\cite{PERETZ2018} proposed a classification of AGNs based on their hardness variability, following a method commonly applied to stellar-mass black holes and X-ray binaries (XRBs). Their study identified four regions (see their Fig.4): Q1 (hard objects that harden as they brighten), Q2 (hard objects that soften as they brighten), Q3 (soft objects that soften as they brighten), and Q4 (soft objects that harden as they brighten). Given the different selection of parameters, a direct comparison between our Figure~\ref{fig:Graph Zone} and the Q1–Q4 classification from \cite{PERETZ2018} is not straightforward. However, the overall distribution of data follows a "harder when brighter" trend, which aligns with Q4, as also found for 1H 0323+342 by \cite{PERETZ2018} based on mean values from their Table~2 (derived from \textit{Swift}/XRT data). Their study did not capture zone 3, which only emerged after 2017.

\begin{figure}[h!]
\centering
\includegraphics[width=0.5\textwidth]{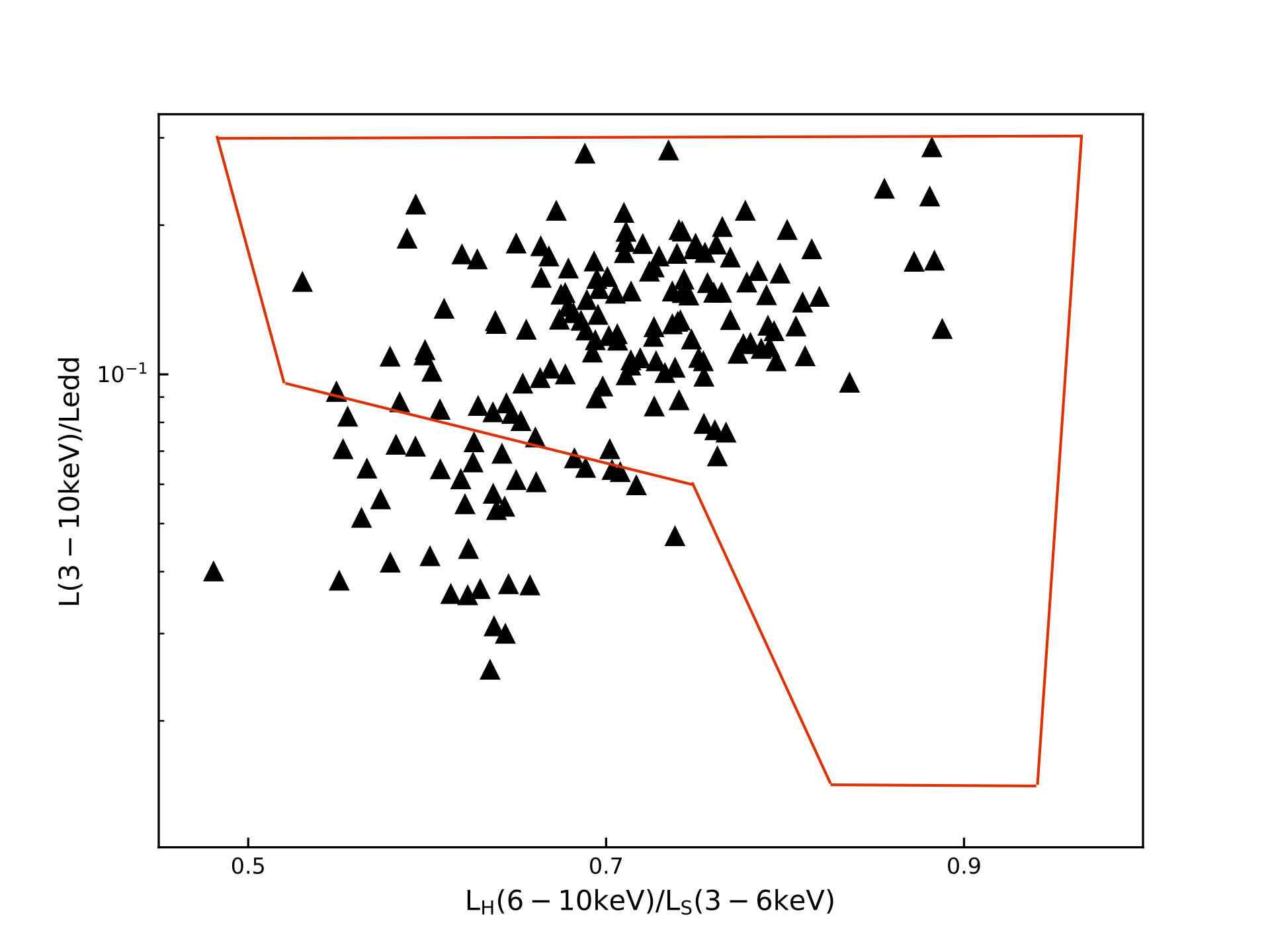}
\caption{Fig.~\ref{fig:Graph Zone} redrawn as an intensity-hardness plot for comparison with XRBs. The dashed red line represents the envelope of XRB regions from \citet[][their Fig.~4]{dunn2010global} and \citet[][their Fig.~1]{MOTTA2021}.}
\label{fig: zonec}
\end{figure}

To facilitate a direct comparison between Fig.~\ref{fig:Graph Zone} and the intensity-hardness plots commonly used for stellar-mass black holes (e.g., \citealt{dunn2010global}, Fig.~4; \citealt{MOTTA2021}, Fig.1), we recalculated the fluxes and replaced the photon index with the hardness ratio. The result, shown in Fig.\ref{fig: zonec}, reveals that while our zone plot exhibits a q-shaped structure similar to that seen in XRBs, it appears as a mirrored version.

In XRBs, spectral transitions occur between a low-hard state, associated with a radio jet and a truncated accretion disk, and a high-soft state, where the jet disappears and the disk extends inward \citep{remillard2006x, plotkin2013x}. In the case of \object{1H 0323+342}, the source alternates between a high-hard state (zone 1), dominated by jet emission, and a high-soft state (zone 2), dominated by the corona. From this high-soft state, the source transitions to a low-soft state (zone 3), where X-ray emission originates solely from the corona.

In the XRB diagram of \cite{dunn2010global}, the empty region corresponds to a low-soft state, where lower luminosities are linked exclusively to hard spectra. Conversely, in Fig.~\ref{fig: zonec}, the empty region corresponds to a low-hard state, while the lowest luminosities are associated only with soft spectra.

This difference arises because XRB intensity-hardness diagrams primarily reflect accretion disk behavior, whereas in AGNs, the X-ray band mainly probes the corona. The lower (UV) frequency peak of disk emission in AGNs, a result of the inverse relationship between disk temperature and black hole mass, contributes to this distinction. Additionally, in jetted AGNs, X-rays can include a significant jet component, such as synchrotron tail emission in BL Lacs or inverse-Compton scattering in quasars and NLS1s. \object{1H 0323+342} stands out due to its strong corona and intermittent jet activity.

As highlighted by \cite{PERETZ2018}, comparing AGNs with XRBs helps identify common physical mechanisms and enhances our understanding of accretion and ejection processes across different mass scales. While AGNs and XRBs can exhibit similar accretion cycles, differences in emission components lead to distinct observational characteristics depending on the source type. In XRBs, jet activity is typically associated with low accretion states. In AGNs, however, the relationship between jet activity and accretion rate is more complex. This difference likely explains why Fig.~\ref{fig: zonec} appears as a mirrored version of the standard XRB diagram. 

\section{Conclusion}
We have conducted a comprehensive multiwavelength analysis of the jetted NLS1 \object{1H 0323+342} to better understand the intricate mechanisms governing disk-corona-jet interactions. Our investigation included a detailed spectral analysis of X-ray data from \emph{Swift}/XRT, photometric analysis of UV and optical data from \emph{Swift}/UVOT, and temporal insights from a \emph{Fermi}/LAT light curve.

We identified three distinct zones in the photon index-flux plot: zone 1, characterized by high fluxes and hard spectra and dominated by jet activity; zone 2, with high fluxes and softer spectra, indicating a weaker jet and a stronger corona; and zone 3, with lower fluxes and soft spectra, where the emission is likely driven solely by the corona. Observations before 2017 primarily show transitions between zones 1 and 2, while post-2017 data reveal transitions between zones 2 and 3, suggesting a cessation or significant weakening of jet activity during this period.

These findings provide strong evidence for intermittent jet activity driven by radiation-pressure instabilities in the accretion disk, consistent with the theoretical framework proposed by \cite{czerny2009accretion}. The observed X-ray flux threshold of $\sim 10^{-11}$ erg cm$^{-2}$ s$^{-1}$ aligns with the accretion rate required to trigger such instabilities.

\object{1H 0323+342} remains a valuable target for studying the interplay between accretion and ejection processes in AGNs. Continued long-term monitoring of this source is crucial for gaining further insights into its evolution.

\section*{Data Availability}
Tables A.1, B.1, C.1, D.1 are only available in electronic form at the CDS via anonymous ftp to 
cdsarc.u-strasbg.fr (130.79.128.5) or via 
\url{http://cdsweb.u-strasbg.fr/cgi-bin/qcat?J/A+A/}.

\bibliography{bibliorosa}{}

\begin{thebibliography}{63}
\expandafter\ifx\csname natexlab\endcsname\relax\def\natexlab#1{#1}\fi

\bibitem[{{Abdo} {et~al.}(2009{\natexlab{a}}){Abdo}, {Ackermann}, \&
  {Ajello}}]{abdo2009bright}
{Abdo}, A.~A., {Ackermann}, M., \& {Ajello}, M. e.~a. 2009{\natexlab{a}}, \apj,
  700, 597

\bibitem[{{Abdo} {et~al.}(2009{\natexlab{b}}){Abdo}, {Ackermann}, \&
  {Ajello}}]{abdo2009radio}
{Abdo}, A.~A., {Ackermann}, M., \& {Ajello}, M. e.~a. 2009{\natexlab{b}}, ApJL,
  707, L142

\bibitem[{Abdollahi {et~al.}(2023)Abdollahi, Ajello, \& Baldini}]{FERMILCR}
Abdollahi, S., Ajello, M., \& Baldini, L. e.~a. 2023, ApJS, 265, 31

\bibitem[{{Angelakis} {et~al.}(2015){Angelakis}, {Fuhrmann}, {Marchili},
  {Foschini}, {Myserlis}, {Karamanavis}, {Komossa}, {Blinov}, {Krichbaum},
  {Sievers}, {et~al.}}]{angelakis2015radio}
{Angelakis}, E., {Fuhrmann}, L., {Marchili}, N., {et~al.} 2015, \aap, 575

\bibitem[{{Berton} {et~al.}(2016){Berton}, {Caccianiga}, \&
  {Foschini}}]{berton2016compact}
{Berton}, M., {Caccianiga}, A., \& {Foschini}, L. e.~a. 2016, \aap, 591, A98

\bibitem[{{Berton}(2019)}]{BERTON2019}
{Berton}, M. e.~a. 2019, \aap, 632, A120

\bibitem[{{Burrows} {et~al.}(2005){Burrows}, {Hill}, {Nousek}, {Kennea},
  {Wells}, {Osborne}, {Abbey}, {Beardmore}, {Mukerjee}, {Short}, {Chincarini},
  {Campana}, {Citterio}, {Moretti}, {Pagani}, {Tagliaferri}, {Giommi},
  {Capalbi}, {Tamburelli}, {Angelini}, {Cusumano}, {Br{\"a}uninger}, {Burkert},
  \& {Hartner}}]{XRT}
{Burrows}, D.~N., {Hill}, J.~E., {Nousek}, J.~A., {et~al.} 2005, \ssr, 120, 165

\bibitem[{{Cardelli} {et~al.}(1989){Cardelli}, {Clayton}, \&
  {Mathis}}]{cardelli1989relationship}
{Cardelli}, J.~A., {Clayton}, G.~C., \& {Mathis}, J.~S. 1989, \apj, 345, 245

\bibitem[{Carpenter \& Ojha(2013)}]{CARPENTER2013}
Carpenter, B. \& Ojha, R. 2013, The Astronomer's Telegram, 5344, 1

\bibitem[{Czerny(2006)}]{CZERNY2006}
Czerny, B. 2006, in Astronomical Society of the Pacific Conference Series, Vol.
  360, AGN Variability from X-Rays to Radio Waves, ed. C.~M. Gaskell, I.~M.
  McHardy, B.~M. Peterson, \& S.~G. Sergeev, 265

\bibitem[{{Czerny} {et~al.}(2009){Czerny}, {Siemiginowska}, {Janiuk},
  {Nikiel-Wroczy{\'n}ski}, \& {Stawarz}}]{czerny2009accretion}
{Czerny}, B., {Siemiginowska}, A., {Janiuk}, A., {Nikiel-Wroczy{\'n}ski}, B.,
  \& {Stawarz}, {\L}. 2009, \apj, 698, 840

\bibitem[{D'Ammando(2020)}]{DAMMANDO2020}
D'Ammando, F. 2020, MNRAS, 496, 2213

\bibitem[{{Doi} {et~al.}(2018){Doi}, {Hada}, {Kino}, {Wajima}, \&
  {Nakahara}}]{doi2018recollimation}
{Doi}, A., {Hada}, K., {Kino}, M., {Wajima}, K., \& {Nakahara}, S. 2018, ApJL,
  857, L6

\bibitem[{{Doi} {et~al.}(2020){Doi}, {Kino}, {Kawakatu}, \& {Hada}}]{DOI2020}
{Doi}, A., {Kino}, M., {Kawakatu}, N., \& {Hada}, K. 2020, \mnras, 496, 1757

\bibitem[{{Dunn} {et~al.}(2010){Dunn}, {Fender}, {K{\"o}rding}, {Belloni}, \&
  {Cabanac}}]{dunn2010global}
{Dunn}, R.~J.~H., {Fender}, R.~P., {K{\"o}rding}, E.~G., {Belloni}, T., \&
  {Cabanac}, C. 2010, \mnras, 403, 61

\bibitem[{{Foschini}(2012)}]{FOSCHINI2012}
{Foschini}, L. 2012, in Proceedings of Nuclei of Seyfert galaxies and QSOs -
  Central engine \& conditions of star formation (Seyfert 2012). 6-8 November,
  10

\bibitem[{{Foschini}(2020)}]{FOSCHINI2020}
{Foschini}, L. 2020, Universe, 6, 136

\bibitem[{{Foschini} {et~al.}(2019){Foschini}, {Ciroi}, \&
  {Berton}}]{foschini2019mapping}
{Foschini}, L., {Ciroi}, S., \& {Berton}, M. e.~a. 2019, Universe, 5, 199

\bibitem[{{Foschini} {et~al.}(2009){Foschini}, {Maraschi}, {Tavecchio},
  {Ghisellini}, {Gliozzi}, \& {Sambruna}}]{FOSCHINI2009}
{Foschini}, L., {Maraschi}, L., {Tavecchio}, F., {et~al.} 2009, Advances in
  Space Research, 43, 889

\bibitem[{{Fuhrmann} {et~al.}(2016){Fuhrmann}, {Karamanavis}, {Komossa},
  {Angelakis}, {Krichbaum}, {Schulz}, {Kreikenbohm}, {Kadler}, {Myserlis},
  {Ros}, {Nestoras}, \& {Zensus}}]{FUHRMANN2016}
{Fuhrmann}, L., {Karamanavis}, V., {Komossa}, S., {et~al.} 2016, Research in
  Astronomy and Astrophysics, 16, 176

\bibitem[{{Gehrels} {et~al.}(2004){Gehrels}, {Chincarini}, {Giommi}, {Mason},
  {Nousek}, {Wells}, {White}, {Barthelmy}, {Burrows}, {Cominsky}, {Hurley},
  {Marshall}, {M{\'e}sz{\'a}ros}, {Roming}, {Angelini}, {Barbier}, {Belloni},
  {Campana}, {Caraveo}, {Chester}, {Citterio}, {Cline}, {Cropper}, {Cummings},
  {Dean}, {Feigelson}, {Fenimore}, {Frail}, {Fruchter}, {Garmire}, {Gendreau},
  {Ghisellini}, {Greiner}, {Hill}, {Hunsberger}, {Krimm}, {Kulkarni}, {Kumar},
  {Lebrun}, {Lloyd-Ronning}, {Markwardt}, {Mattson}, {Mushotzky}, {Norris},
  {Osborne}, {Paczynski}, {Palmer}, {Park}, {Parsons}, {Paul}, {Rees},
  {Reynolds}, {Rhoads}, {Sasseen}, {Schaefer}, {Short}, {Smale}, {Smith},
  {Stella}, {Tagliaferri}, {Takahashi}, {Tashiro}, {Townsley}, {Tueller},
  {Turner}, {Vietri}, {Voges}, {Ward}, {Willingale}, {Zerbi}, \&
  {Zhang}}]{gehrels2004swift}
{Gehrels}, N., {Chincarini}, G., {Giommi}, P., {et~al.} 2004, \apj, 611, 1005

\bibitem[{{Ghosh} {et~al.}(2018){Ghosh}, {Dewangan}, {Mallick}, \&
  {Raychaudhuri}}]{GHOSH2018}
{Ghosh}, R., {Dewangan}, G.~C., {Mallick}, L., \& {Raychaudhuri}, B. 2018,
  \mnras, 479, 2464

\bibitem[{{Goodrich}(1989)}]{GOODRICH1989}
{Goodrich}, R.~W. 1989, \apj, 342, 224

\bibitem[{{Grupe} \& {Mathur}(2004)}]{grupe2004mbh}
{Grupe}, D. \& {Mathur}, S. 2004, The Astrophysical Journal Letters, 606, L41

\bibitem[{{Haardt} \& {Maraschi}(1991)}]{HAARDT1991}
{Haardt}, F. \& {Maraschi}, L. 1991, \apjl, 380, L51

\bibitem[{{Haardt} \& {Maraschi}(1993)}]{HAARDT1993}
{Haardt}, F. \& {Maraschi}, L. 1993, \apj, 413, 507

\bibitem[{{Hada} {et~al.}(2018){Hada}, {Doi}, {Wajima}, {D'Ammando}, {Orienti},
  {Giroletti}, {Giovannini}, {Nakamura}, \& {Asada}}]{HADA2018}
{Hada}, K., {Doi}, A., {Wajima}, K., {et~al.} 2018, \apj, 860, 141

\bibitem[{{HI4PI Collaboration} {et~al.}(2016){HI4PI Collaboration}, {Ben
  Bekhti}, {Fl{\"o}er}, {Keller}, {Kerp}, {Lenz}, {Winkel}, {Bailin},
  {Calabretta}, {Dedes}, {Ford}, {Gibson}, {Haud}, {Janowiecki}, {Kalberla},
  {Lockman}, {McClure-Griffiths}, {Murphy}, {Nakanishi}, {Pisano}, \&
  {Staveley-Smith}}]{GALACTICNH}
{HI4PI Collaboration}, {Ben Bekhti}, N., {Fl{\"o}er}, L., {et~al.} 2016, \aap,
  594, A116

\bibitem[{Iso {et~al.}(2016)Iso, Ebisawa, Sameshima, Mizumoto, Miyakawa, Inoue,
  \& Yamasaki}]{ISO2016}
Iso, N., Ebisawa, K., Sameshima, H., {et~al.} 2016, PASJ, 68, S27

\bibitem[{{Itoh} {et~al.}(2014){Itoh}, {Tanaka}, {Akitaya}, {Uemura},
  {Fukazawa}, {Inoue}, {Doi}, {Arai}, {Hanayama}, {Hashimoto}, {Hayashi},
  {Izumiura}, {Kanda}, {Kawabata}, {Kawaguchi}, {Kawai}, {Kinugasa}, {Kuroda},
  {Miyaji}, {Moritani}, {Morokuma}, {Murata}, {Nagayama}, {Oasa}, {Ohshima},
  {Ohsugi}, {Saito}, {Sakata}, {Sasada}, {Sekiguchi}, {Takagi}, {Takahashi},
  {Takaki}, {Ui}, {Watanabe}, {Yamanaka}, {Yamashita}, \& {Yoshida}}]{ITOH2014}
{Itoh}, R., {Tanaka}, Y.~T., {Akitaya}, H., {et~al.} 2014, \pasj, 66, 108

\bibitem[{{Kynoch}(2018)}]{KYNOCH2018}
{Kynoch}, D. e.~a. 2018, \mnras, 475, 404

\bibitem[{{Landt} {et~al.}(2017){Landt}, {Ward}, {Balokovi{\'c}}, {Kynoch},
  {Storchi-Bergmann}, {Boisson}, {Done}, {Schimoia}, \&
  {Stern}}]{landt2017black}
{Landt}, H., {Ward}, M.~J., {Balokovi{\'c}}, M., {et~al.} 2017, \mnras, 464,
  2565

\bibitem[{Leighly(1999)}]{Leighly1999}
Leighly, K.~M. 1999, ApJS, 125, 317

\bibitem[{{Le{\'o}n Tavares} {et~al.}(2014){Le{\'o}n Tavares}, {Kotilainen},
  {Chavushyan}, {A{\~n}orve}, {Puerari}, {Cruz-Gonz{\'a}lez},
  {Pati{\~n}o-Alvarez}, {Ant{\'o}n}, {Carrami{\~n}ana}, {Carrasco}, {Guichard},
  {Karhunen}, {Olgu{\'\i}n-Iglesias}, {Sanghvi}, \& {Valdes}}]{LEONTAV2014}
{Le{\'o}n Tavares}, J., {Kotilainen}, J., {Chavushyan}, V., {et~al.} 2014,
  \apj, 795, 58

\bibitem[{{Luashvili}(2023)}]{LUASHVILI2023}
{Luashvili}, A. e.~a. 2023, \mnras, 523, 404

\bibitem[{Mattox {et~al.}(1996)Mattox, Bertsch, Chiang, Dingus, Digel,
  Esposito, Fierro, Hartman, Hunter, Kanbach, Kniffen, Lin, Macomb,
  Mayer-Hasselwander, Michelson, von Montigny, Mukherjee, Nolan, Ramanamurthy,
  Schneid, Sreekumar, Thompson, \& Willis}]{MATTOX1996}
Mattox, J.~R., Bertsch, D.~L., Chiang, J., {et~al.} 1996, ApJ, 461, 396

\bibitem[{Motta {et~al.}(2021)Motta, Rodriguez, Jourdain, {Del Santo},
  Belanger, Cangemi, Grinberg, Kajava, Kuulkers, Malzac, Pottschmidt, Roques,
  Sánchez-Fernández, \& Wilms}]{MOTTA2021}
Motta, S., Rodriguez, J., Jourdain, E., {et~al.} 2021, New Astronomy Reviews,
  93, 101618

\bibitem[{{Mundo}(2020)}]{MUNDO2020}
{Mundo}, S.~A. e.~a. 2020, \mnras, 496, 2922

\bibitem[{{Netzer}(2019)}]{netzer2019bolometric}
{Netzer}, H. 2019, MNRAS, 488, 5185

\bibitem[{{Olgu{\'\i}n-Iglesias} {et~al.}(2020){Olgu{\'\i}n-Iglesias},
  {Kotilainen}, \& {Chavushyan}}]{OLGUIN2020}
{Olgu{\'\i}n-Iglesias}, A., {Kotilainen}, J., \& {Chavushyan}, V. 2020, \mnras,
  492, 1450

\bibitem[{{Osterbrock} \& {Pogge}(1985)}]{osterbrock1985spectra}
{Osterbrock}, D.~E. \& {Pogge}, R.~W. 1985, \apj, 297, 166

\bibitem[{{Paliya} {et~al.}(2014){Paliya}, {Sahayanathan}, {Parker}, {Fabian},
  {Stalin}, {Anjum}, \& {Pandey}}]{paliya2014peculiar}
{Paliya}, V.~S., {Sahayanathan}, S., {Parker}, M.~L., {et~al.} 2014, \apj, 789,
  143

\bibitem[{Paliya {et~al.}(2015)Paliya, Stalin, \& Ravikumar}]{PALIYA2015}
Paliya, V.~S., Stalin, C.~S., \& Ravikumar, C.~D. 2015, AJ, 149, 41

\bibitem[{{Peretz} \& {Behar}(2018)}]{PERETZ2018}
{Peretz}, U. \& {Behar}, E. 2018, \mnras, 481, 3563

\bibitem[{{Peterson} \& {Dalla Bont{\`a}}(2018)}]{PETERSON2018}
{Peterson}, B. \& {Dalla Bont{\`a}}, E. 2018, in Revisiting Narrow-Line Seyfert
  1 Galaxies and their Place in the Universe, 8

\bibitem[{{Peterson}(2011)}]{PETERSON2011}
{Peterson}, B.~M. 2011, in Narrow-Line Seyfert 1 Galaxies and their Place in
  the Universe, ed. L.~{Foschini}, M.~{Colpi}, L.~{Gallo}, D.~{Grupe},
  S.~{Komossa}, K.~{Leighly}, \& S.~{Mathur}, 32

\bibitem[{{Plotkin} {et~al.}(2013){Plotkin}, {Gallo}, \&
  {Jonker}}]{plotkin2013x}
{Plotkin}, R.~M., {Gallo}, E., \& {Jonker}, P.~G. 2013, \apj, 773, 59

\bibitem[{{Remillard} {et~al.}(1993){Remillard}, {Bradt}, {Brissenden},
  {Buckley}, {Roberts}, {Schwartz}, {Stroozas}, \&
  {Tuohy}}]{remillard1993twenty}
{Remillard}, R.~A., {Bradt}, H.~V.~D., {Brissenden}, R.~J.~V., {et~al.} 1993,
  AJ, 105, 2079

\bibitem[{{Remillard} \& {McClintock}(2006)}]{remillard2006x}
{Remillard}, R.~A. \& {McClintock}, J.~E. 2006, Annual Review of Astronomy and
  Astrophysics, 44, 49

\bibitem[{{Roming} {et~al.}(2005){Roming}, {Kennedy}, {Mason}, {Nousek}, {Ahr},
  {Bingham}, {Broos}, {Carter}, {Hancock}, {Huckle}, {Hunsberger}, {Kawakami},
  {Killough}, {Koch}, {McLelland}, {Smith}, {Smith}, {Soto}, {Boyd},
  {Breeveld}, {Holland}, {Ivanushkina}, {Pryzby}, {Still}, \& {Stock}}]{UVOT}
{Roming}, P. W.~A., {Kennedy}, T.~E., {Mason}, K.~O., {et~al.} 2005, \ssr, 120,
  95

\bibitem[{Rybicki \& Lightman(1979)}]{RYBLIG}
Rybicki, G.~B. \& Lightman, A.~P. 1979, Radiative processes in astrophysics
  (New York, USA: J. Wiley\& Sons)

\bibitem[{{Schlafly} \& {Finkbeiner}(2011)}]{schlafly2011measuring}
{Schlafly}, E.~F. \& {Finkbeiner}, D.~P. 2011, ApJ, 737, 103

\bibitem[{{Tibolla} {et~al.}(2013){Tibolla}, {Kaufmann}, {Foschini},
  {Mannheim}, {Zhang}, {Li}, {Angelakis}, {Fuhrmann}, {H{\"a}usner}, {Kania},
  {Els{\"a}sser}, {Kreikenbohm}, {Schulz}, \& {Kadler}}]{TIBOLLA2013}
{Tibolla}, O., {Kaufmann}, S., {Foschini}, L., {et~al.} 2013, in International
  Cosmic Ray Conference, Vol.~33, International Cosmic Ray Conference, 2748

\bibitem[{{Turner} {et~al.}(2022){Turner}, {Miller}, {Maune}, \&
  {Eggen}}]{TURNER2022}
{Turner}, C.~S., {Miller}, H.~R., {Maune}, J.~D., \& {Eggen}, J.~R. 2022,
  \mnras, 517, 3257

\bibitem[{Waddell \& Gallo(2020)}]{WADDELL2020}
Waddell, S.~G.~H. \& Gallo, L.~C. 2020, MNRAS, 498, 5207

\bibitem[{{Wajima} {et~al.}(2014){Wajima}, {Fujisawa}, {Hayashida}, {Isobe},
  {Ishida}, \& {Yonekura}}]{wajima2014short}
{Wajima}, K., {Fujisawa}, K., {Hayashida}, M., {et~al.} 2014, ApJ, 781, 75

\bibitem[{Walton {et~al.}(2013)Walton, Nardini, Fabian, Gallo, \&
  Reis}]{WALTON2013}
Walton, D.~J., Nardini, E., Fabian, A.~C., Gallo, L.~C., \& Reis, R.~C. 2013,
  MNRAS, 428, 2901

\bibitem[{{Wang} {et~al.}(2016){Wang}, {Du}, {Hu}, {Bai}, {Wang}, {Yi}, {Wang},
  {Zhang}, {Xin}, {Lun}, {Chang}, \& {Fan}}]{wang2016reverberation}
{Wang}, F., {Du}, P., {Hu}, C., {et~al.} 2016, \apj, 824, 149

\bibitem[{{Wang} {et~al.}(2017){Wang}, {Xiong}, {Bai}, {Li}, \&
  {Wang}}]{WANG2017}
{Wang}, F., {Xiong}, D.-R., {Bai}, J.-M., {Li}, S.-K., \& {Wang}, J.-G. 2017,
  Research in Astronomy and Astrophysics, 17, 068

\bibitem[{{Wood} {et~al.}(1984){Wood}, {Meekins}, {Yentis}, {Smathers},
  {McNutt}, {Bleach}, {Byram}, {Chupp}, {Friedman}, \& {Meidav}}]{wood1984heao}
{Wood}, K.~S., {Meekins}, J.~F., {Yentis}, D.~J., {et~al.} 1984, \apjs, 56, 507

\bibitem[{{Yao} \& {Komossa}(2023)}]{YAO2023}
{Yao}, S. \& {Komossa}, S. 2023, \mnras, 523, 441

\bibitem[{{Yao} {et~al.}(2015){Yao}, {Yuan}, {Komossa}, {Grupe}, {Fuhrmann}, \&
  {Liu}}]{YAO2015}
{Yao}, S., {Yuan}, W., {Komossa}, S., {et~al.} 2015, \aj, 150, 23

\bibitem[{Zhou {et~al.}(2007)Zhou, Wang, Yuan, \& et~al.}]{Zhou2007}
Zhou, H., Wang, T., Yuan, W., \& et~al. 2007, The Astrophysical Journal, 658,
  L13

\end{thebibliography}
\bibliographystyle{aa}

\end{document}